\documentclass[conference]{IEEEtran}
\usepackage{cite}
\ifCLASSINFOpdf
\else
\fi
\usepackage{graphicx}
\usepackage{algorithmic}
\usepackage{array}

\begin{document}

% renew command to make font in verbatim smaller
\makeatletter
\renewcommand\verbatim@font{\small\normalfont\ttfamily}
\makeatother

\renewcommand\IEEEkeywordsname{Keywords}
\title{Zero Knowledge Proof for \\ Multiple Sequence Alignment}

\iftrue
\author{\IEEEauthorblockN{Worasait Suwannik}
\IEEEauthorblockA{Department of Computer Science\\
Kasetsart University\\
Bangkok, Thailand\\
worasait.suwannik@gmail.com}
}
\fi

\iffalse
\author{\IEEEauthorblockN{}
\IEEEauthorblockA{\\
\\
\\
}
}
\fi

\maketitle

\begin{abstract}
Multiple sequence alignment (MSA) is a fundamental algorithm in bioinformatics.
In a situation when the alignment might need to be protected while revealing the other information such the input sequences and the alignment score, zero knowledge proof can be used.
In this paper, a validator checks the consistency between the input sequence and the alignment, and between the alignment and the alignment score.
The validator is written in Circom language which will be compile into a circuit.
Using a zero knowledge prove system called zkSNARK, a cryptographic proof is generates for the circuit and its input.
This proof demonstrates that all inputs are consistent without revealing the actual alignment.
\end{abstract}


\section{Introduction}
Multiple sequence alignment (MSA) is a fundamental algorithm in bioinformatics that addresses the problem of comparing and optimally aligning multiple biological sequences, often DNA, RNA, or proteins. From a computer science perspective, an MSA algorithm can be viewed as a string alignment problem where the input is a list of strings, which represent biological sequences and the output is another list of strings, which represent the alignment. The objective of the algorithm is to find an alignment that maximizes the similarities between the sequences while accounting for insertions, deletions, and mutations that may have occurred.  

\begin{figure}
    \centering
    \includegraphics[width=1\linewidth]{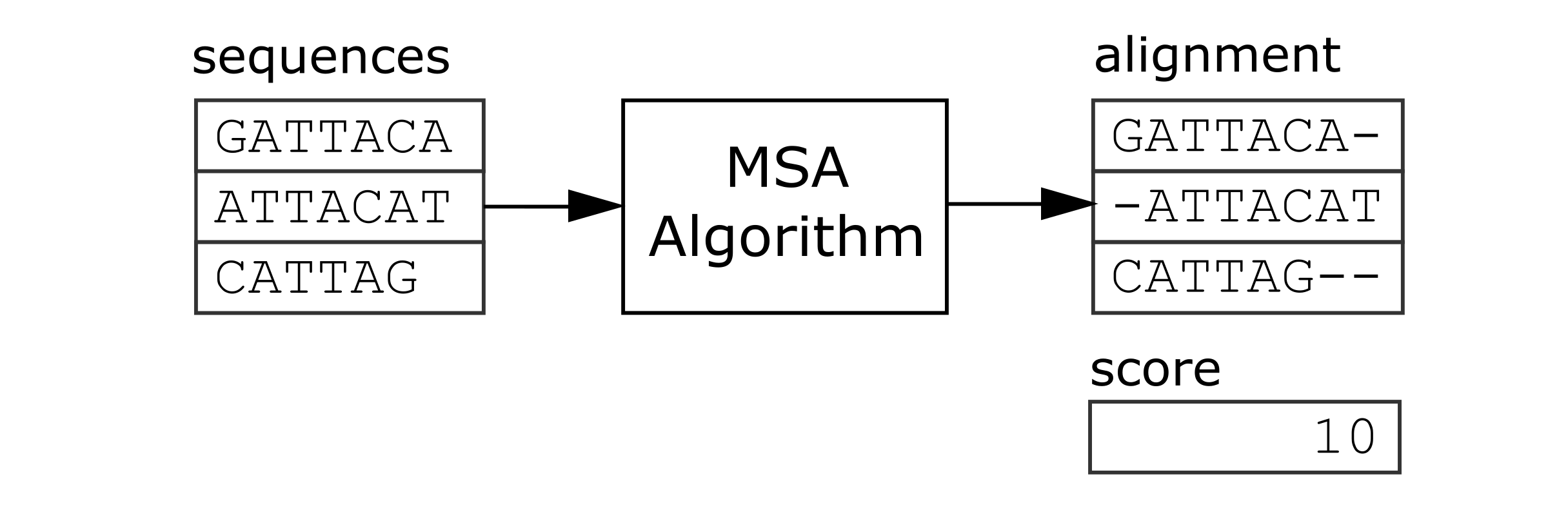}
    \caption{Input and output of multiple sequence alignment}
    \label{fig:msa}
\end{figure}

Example of MSA is shown in Figure \ref{fig:msa}.  The input of MSA are sequences.  Its output is the alignment that has a certain score.  The time complexity for optimally aligning $k$ sequences of length $n$ using original dynamic programming is $O(n^k 2^k)$ \cite{newDP4MSA}. Therefore, heuristics, which are not optimally guarantee, are normally used to align multiple sequences.

This alignment allows researchers to identify conserved regions – crucial for protein function or regulatory elements – and infer evolutionary relationships between the sequences, to understand biological processes.  It can also help in developing new drugs by identifying conserved regions in protein sequences that may serve as potential drug targets.

In commercially driven bioinformatics fields, a challenge lies in balancing transparency, crucial for scientific progress, with confidentiality to protect a competitive edge. Within the context of MSA, withholding the exact MSA results while disclosing the alignment score might be an approach to balancing transparency and confidentiality. This score hints at the analysis quality and allows for an assessment of its validity. This approach fosters collaboration and future research by enabling some level of scientific exchange, all while safeguarding commercially sensitive sequence information. 

Various approaches are used to protect intellectual property in bioinformatics including legal measures, such as copyright patent, and other law tactics \cite{ip_strategry_bioinfo}; to using technical measures.   Techniques to control the threat includes: secure multi-party computation, cryptographic hash function, homomorphic encryption, a trust execution environment, blockchain technology, and a filter of privacy-sensitive information \cite{privacy_preserving_seq_alignment}.

A promising approach to addressing the challenge of balancing transparency with confidentiality is the use of a cryptographic method known as zero-knowledge proofs (ZKP) \cite{Goldwasser_interactive_proof}. ZKP allow a researcher to prove the validity of their MSA result, without revealing the alignment detail. The researcher (prover) can mathematically convince a verifier (e.g., another scientist or company) that their alignment score is accurate, without disclosing the underlying alignments that hold commercial value. This approach fosters collaboration and scientific progress while safeguarding sensitive data, potentially revolutionizing how bioinformatics handles confidential information.

\begin{figure}
    \centering
    \includegraphics[width=1\linewidth]{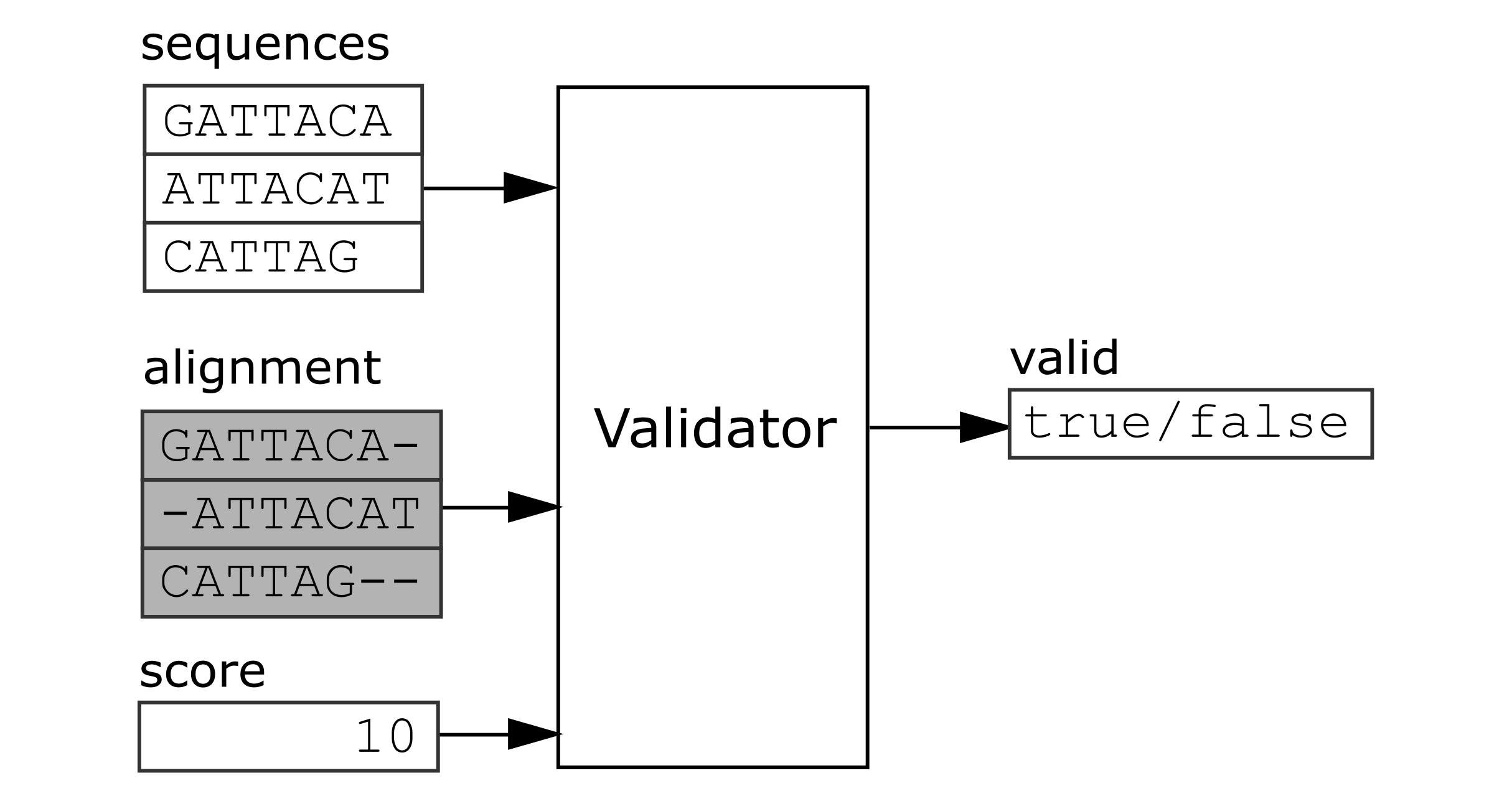}
    \caption{Validator.  Private input has gray background.}
    \label{fig:validator}
\end{figure}

This paper proposed a zero knowledge proof for multiple sequence alignment.  We wrote a program that validates input and output of MSA using Circom language (Figure \ref{fig:validator}).  Circom \cite{Circom} is a domain specific language to define circuits that perform calculations, with the special property that proofs can be generated showing the validity of the results without revealing the underlying some or all the inputs. The compiled code can be used with other tools such as snark.js to generate and verify a proof.

% compare this sudoku or Hash preimage

\section{Methodology}
Validating the input and the output of MSA involves two parts.  The first is to checks if the alignment reflects the score assigned by the MSA algorithm.  The second is to check whether if the aligned sequences match their original forms.  For example, the sequence GATTACA matches the alignment GATTACA- or GAT-TA-CA or --GATTA-CA- but does not match CATTACA. 

\begin{figure}
    \centering
    \includegraphics[width=1\linewidth]{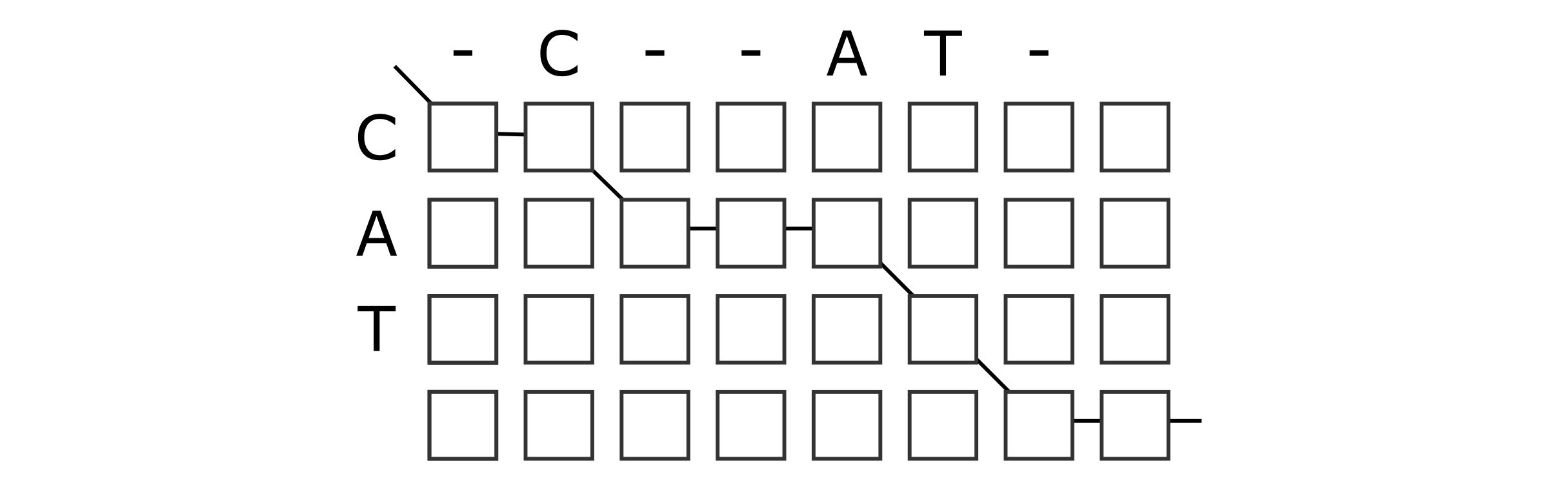}
    \caption{2D array of components for checking the consistency of the alignment and the input sequence}
    \label{fig:routing}
\end{figure}
\subsection{Consistency of the alignment and the score}

Circom stands for circuit compiler.  In Circom, a programmer define circuits that perform calculations. These circuits have inputs , outputs, and internal components that connect to each other.  A component is created from a blueprint called a template.  A component has input and output signals and can connect to other components.  

The following code is a template for MSA scoring system.  In MSAs, sequences are compared to identify regions of similarity. The scoring system assigns points for matches (alignment), mismatches (different characters), and gaps (insertions/deletions). This Circom template represents one of a scoring systems in bioinformatics, where matches score 1, mismatches score -1, and gaps score -1.  IsEqual, AND, OR, NOT are from a circomlib library.  Note that the signal in Circom is not just 0 or 1 as in a digital circuit.  The value of the signal is in the range of 0 to 21888242871839275222246405745257275088548364400416 034343698204186575808495616.  The value can represent minus.  For example, -1 is 21888242871839275222246405745257275088548364400416 034343698204186575808495616. 

\begin{verbatim}
template scoring_system() {
  signal input x[2];
  signal output y;

  signal xeq  <== IsEqual()(x);   
  signal xneq <== NOT()(xeq);
  signal gap  <== OR()(IsEqual()([x[0],0]), 
                       IsEqual()([x[1],0]));
  signal bgap <== AND()(xeq,gap);   
  signal ngap <== NOT()(gap);         
  signal eq_ngap <== AND()(xeq, ngap);

  y <== (eq_ngap * 1) +  // match
        (xneq * -1)   +  // mismatch
        (bgap * -1);     // two gaps
}
\end{verbatim}

The symbol \textless == means assignment to a signal and defining a constraint.  The scoring\textunderscore system component has 9 constraints.  The proof is verified by checking if the constraints hold true.

To calculate a score of a pair of alignment, the scoring\textunderscore system component is created for each element in the alignment.

\begin{verbatim}
template pair_score(aln_len) {
  signal input aln[2][aln_len];
  signal output score;
  component c[seq_len];
  signal s[seq_len];

  for (var i = 0; i < aln_len; i++) {
    c[i] = scoring_system();
    c[i].x[0] <== aln[0][i];
    c[i].x[1] <== aln[1][i];
    s[i] <== (i == 0 ? 0 : s[i-1]) + 
             c[i].y;
  }
  score <== s[aln_len - 1];
}
\end{verbatim}

To evaluate MSA score, every pair of alignment has to be evaluated.  A component pair\textunderscore score is created for each pair of alignment.  Then the sum of scores is sent as an output of a template msa\textunderscore score, which output is compared with the score from user input.  If both scores are equal, the component check\textunderscore aln\textunderscore score will output 1. Otherwise, it will output 0.

\begin{verbatim}
template msa_score(nseq, aln_len) {
  signal input aln[nseq][aln_len];
  signal output score;
  ...
}

template check_aln_score(nseq, aln_len) {
  signal input aln[nseq][aln_len];
  signal input score;
  signal output y;
  signal s <== msa_score(nseq, aln_len)(aln);
  y <== IsEqual()([s, score]);
}
\end{verbatim}

\subsection{Consistency of the alignment and the input sequences}

Another part of the circuit checks the consistency of the sequences and the alignment. For each sequence and alignment pair, a 2-dimensional array of components is created. An enable signal is sent to a component at the top left corner.  A component forwards an enable signal to another component in the array if the sequence and the alignment are consistent.  Example of signal routing is shown in Figure \ref{fig:routing}.  

There are two types of components in the 2D array. The type T2 is located at the last row of the array.  An enable signal that reaches this row indicates a complete match of all letters in the sequence and the alignment. The circuit in this row verifies the absence of any remaining letters (excluding gaps) in the alignment.  The following is the template for the component in the last row. The input e enables the component.  The input c is an element in an alignment string.

\begin{verbatim}  
template T2() {
  signal input e, c;  
  signal output ee;   
  signal e1 <== IsEqual()([e,1]); 
  signal c0 <== IsEqual()([c,0]);  
  ee <== e1 * c0;
}
\end{verbatim}  

Another type of component (T1) is also responsible for routing the enable signal in the circuit.  In 2D array, the components of this type is placed on top of the T2 row. Similar to T2, the input e enables the component and the input c is an element in an alignment string.  The input r is an element in a sequence.  This component enables the right component (the East direction) if input c is a gap and r is a letter.  It enables the bottom component (the South direction) if there is no more letter in the sequence.  It enables the bottom right (the South East direction) if letters from the sequence and the alignment are the same.  It enables at most one neighbor component.  The following is the template for T1.

\begin{verbatim}  
template T1() {
  signal input e, r, c;  // enabled, row, col
  signal output es, ese, ee;  
  signal e1  <== IsEqual()([e,1]); 
  signal rc  <== IsEqual()([c,r]);   
  signal c0  <== IsEqual()([c,0]);  
  signal r0  <== IsEqual()([r,0]);  
  signal nr0 <== NOT()(r0);  

  signal t1 <== rc * nr0;
  signal t2 <== c0 * nr0;

  es  <== e1 * r0;  
  ese <== e1 * t1;
  ee  <== e1 * t2;
}
\end{verbatim}  

\subsection{Main Component}

The main component receives the output from the previous two subcomponents.  It outputs 1 if both are checked.  There are 3 inputs to the component as shown in Figure \ref{fig:validator}.  The last line of the code indicates that only two of them (i.e., the sequences and the score) are public input.  The alignment is hidden.

\begin{verbatim}    
template Main(nseq, seq_len, aln_len) {
  signal input seq[nseq][seq_len];
  signal input aln[nseq][aln_len];
  signal input score;
  signal output y;

  signal aln_seq   <== 
    check_aln_seq(nseq, seq_len, aln_len)
                 (seq, aln);
  signal aln_score <== 
    check_aln_score(nseq, aln_len)
                   (aln, score);
  y <== AND()(aln_score, aln_seq);
}

component main {public [seq, score]} = 
          Main(4,8,11);
\end{verbatim}

\subsection{Generating and Verifying a Proof}

Circom acts as a high-level language for writing a verifiable circuit, which encodes the logic of a computation. zkSNARKs (Zero Knowledge Succinct Non-interactive Arguments of Knowledge) creates cryptographic proofs that the circuit completed a specific computation without revealing the actual inputs used. Groth16 is an efficient zkSNARK construction often used with Circom \cite{Groth16}. 

\begin{figure}
    \centering
    \includegraphics[width=1\linewidth]{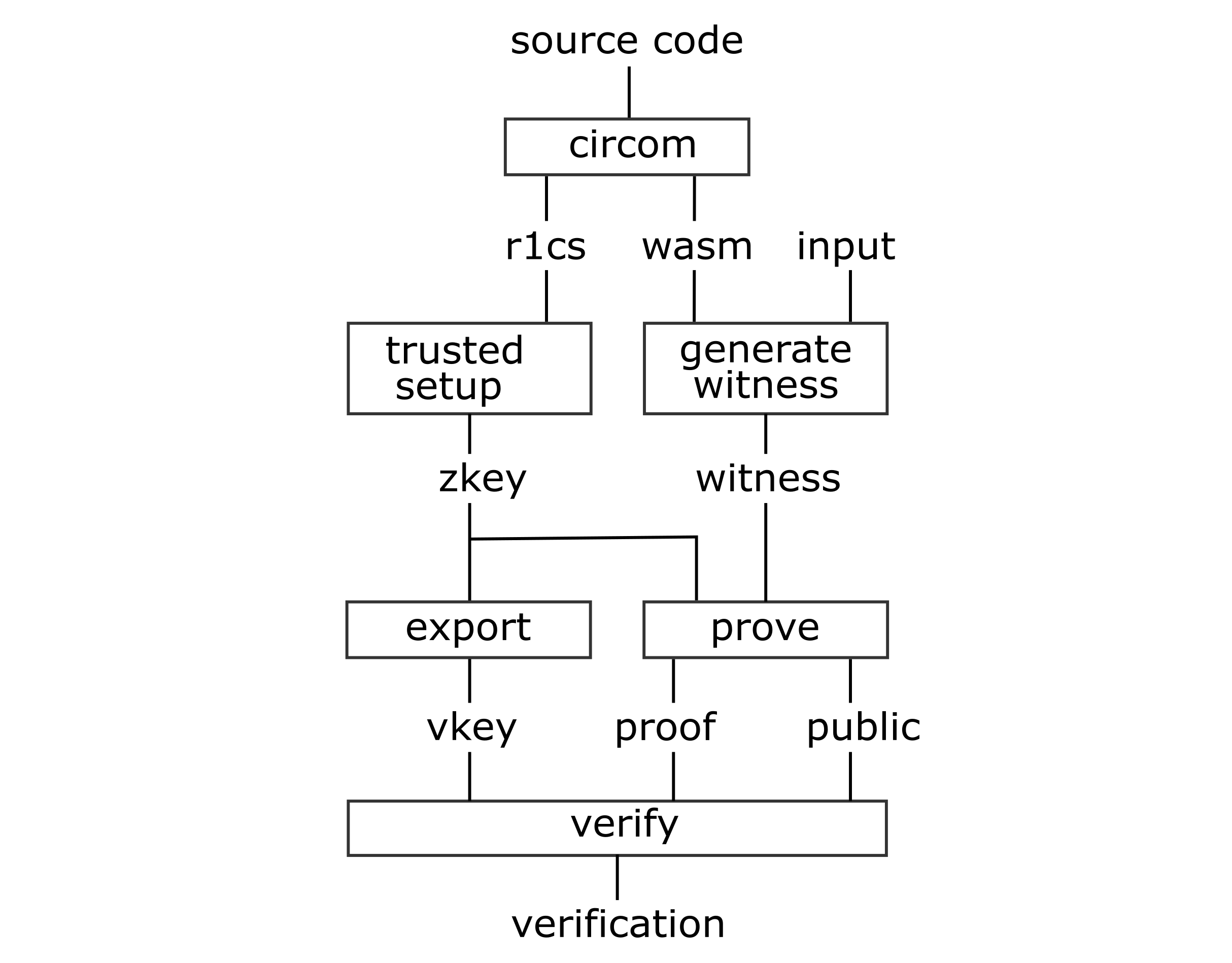}
    \caption{Zero knowledge proof using Circom}
    \label{fig:zkprove}
\end{figure}

The whole process of zero knowledge proof using Circom and snarkjs (i.e., a zkSNARK implementation in JavaScript) is shown in Figure \ref{fig:zkprove}.  A Circom source code is compiled.  The result is a circuit and a web assembly (wasm) file.  After that the web assembly and input files are sent the circuit to generate a witness.  A witness is values of signals and variables in the circuit.  The circuit is sent to a trust set up generate a zkey.  During this set up phase, a random number is created and then should immediately be destroyed to guarantees that only proofs based on valid witnesses can be created.  During this step, multiple parties contribute randomness to ensure no single party has complete control.   The prover uses the zkey and the witness to generate a proof and a public output.  The verifier uses them and a verification key to verify the proof. If there is some discrepancy, such as changing a public input, the verification will fail.

\section{Result}

Circom compiler outputs circuits information including: number of template instances, non-linear constraints, linear constraints, public inputs/outputs, private inputs/outputs, 
wires, and labels.
The number of constraints in a Circom circuit is important for zero-knowledge prove because it affects the efficiency and security of the proof. Under-constrained circuits may make the proof vulnerable to attacks, while over-constrained circuit may make the proof slow to generate or verify.
Experimentation with various input sizes is conducted to observe their impact on the number of constraints.
As shown in Table \ref{tab:result}, the number of non-linear constraints is $O(N\_ Seq \times Seq\_  Len \times Aln\_  Len) $.

\begin{table}
    \centering
    \caption{Number of constraints at various input parameters.  N\textunderscore Seq is the number of sequences.  Seq\textunderscore Len is the length of sequences.  Aln\textunderscore Len is the length of alignment.}
    \label{tab:result}
    \begin{tabular}{rrrr}
        N\textunderscore Seq & Seq\textunderscore  Len & Aln\textunderscore  Len & Number of Constraints \\
        10 & 10 & 10 & 18552  \\
        10 & 10 & 100 & 181002  \\
        10 & 100 & 100 & 1355502  \\
        100 & 100 & 100 & 17605002  \\
        100 & 100 & 200 & 35160002  \\
    \end{tabular}
\end{table}

\section{Discussion}

The parameter in Table \ref{tab:result} aims for better understanding of relationship between the input and the number of the constraints. For the real usage of MSA, the number of sequences can vary depending on the specific goals of the analysis and the computational capabilities of the system.  In some cases, researchers might focus on a small set of closely related sequences to understand conservation within a particular protein family, while in other cases, they might analyze a larger dataset spanning multiple species to identify broadly conserved features. Our design cannot handle very large dataset because the number of constraint will be too large.  For example, the compiler version 2.1.8 stopped without reporting the number of constraints with the parameter 1000 100 150 on 16GB Apple M2 chip.  

Checking the correspondences of inputs using hardware description paradigm is more difficult that writing a high level language code. Calculating score between of two alignment strings using Python requires only 8 line of code.  While checking the correspondence between the sequence and the alignment is one liner using a built-in method. In contrast, equality checking between two signal in circomlib uses 15 lines of Circom code.

\begin{figure}
    \centering
    \includegraphics[width=1\linewidth]{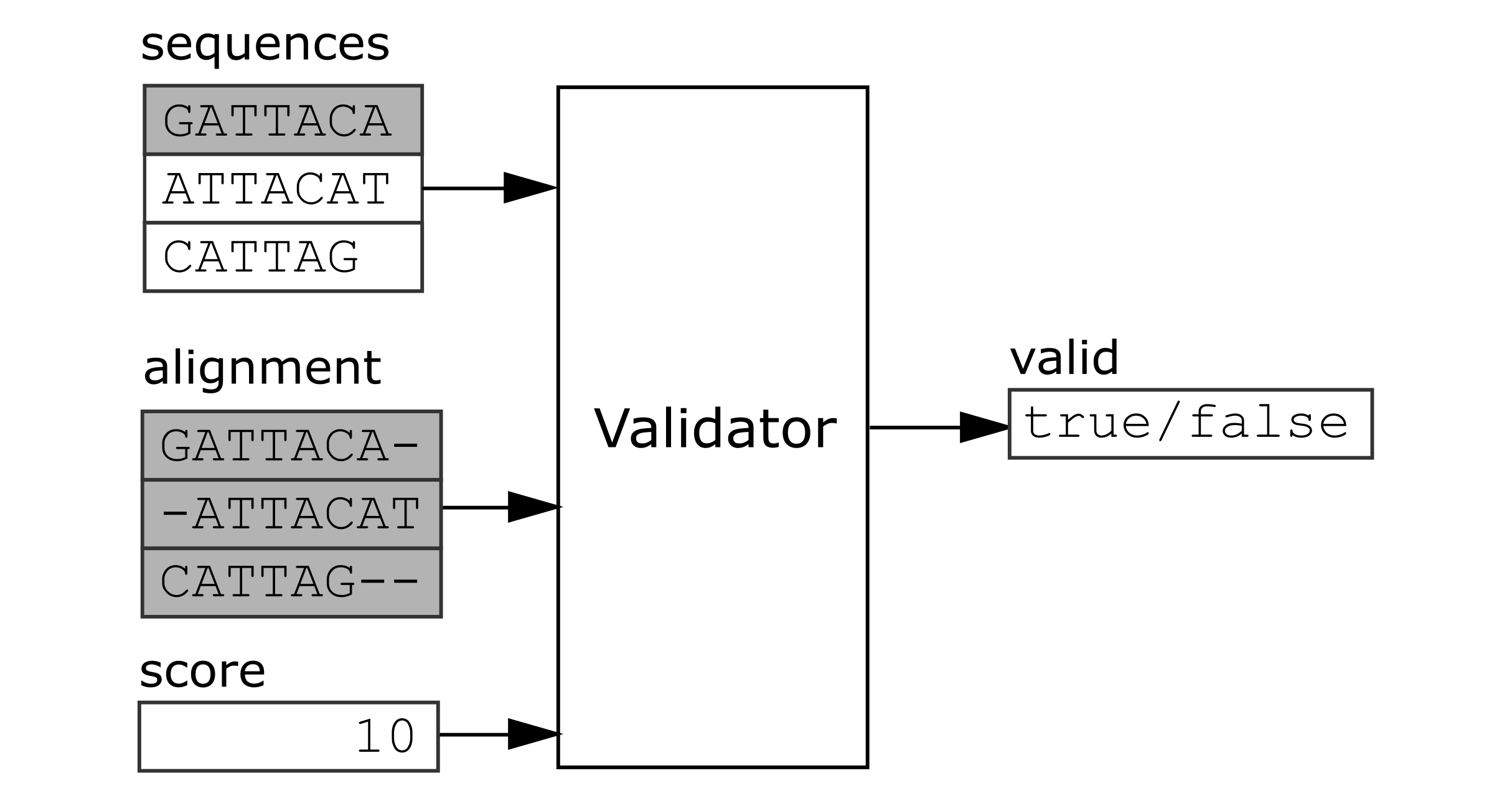}
    \caption{Another use case of zero knowledge prove for MSA.  Private inputs have gray background.}
    \label{fig:validator2}
\end{figure}

For other use cases, this Circom source code can be modified to hide the alignment and one input sequence while revealing the other input sequences and score (Figure \ref{fig:validator2}).  In addition, it can be used to show one input sequence and score while hiding the other input sequences, the alignment, and the score.

\section{Conclusion}
This work presents the design of circuit for zero knowledge proof for multiple sequence alignment.  Our design uses basic operations such as comparators and basic boolean logic from circomlib library.  Even though our construction can validate the sequence, the alignment, and the score, the number of constraints is very high.  By optimizing the circuit to reduce constraints in the future, we aim to enable the proving system to handle larger and more complex sequence alignments.

\section*{Acknowledgment}

The author would like to thank Somrak Numnark for valuable discussion and for comments that greatly improved the manuscript.

\bibliographystyle{IEEEtran}
\bibliography{refs}

\end{document}